\documentclass[cameraready]{Interspeech}

\title{EffVOC: Low-Delay Efficient Speech Waveform Reconstruction \\ from Spectral Representations Without Phase}

\author[affiliation={1}]{Renzheng}{Shi}
\author[affiliation={2}]{Simon}{Welker}
\author[affiliation={2}]{Timo}{Gerkmann}
\author[affiliation={1}]{Tim}{Fingscheidt}

\address{
    $^1$ Institute for Communications Technology, TU Braunschweig, Braunschweig, Germany \\
    $^2$ Signal Processing Group, Universität Hamburg, Hamburg, Germany
}

\email{\{re.shi, t.fingscheidt\}@tu-bs.de, \{simon.welker, timo.gerkmann\}@uni-hamburg.de}


\usepackage{comment}

\usepackage{cite}
\usepackage{amsmath,amssymb,amsfonts}
\usepackage{algorithmic}
\usepackage{graphicx}
\usepackage{textcomp}
\usepackage{xcolor}

\usepackage[mode=buildnew]{standalone}
\usepackage{subfigure}
\usepackage{multirow}
\usepackage{makecell}
\usepackage{bbding}
\usepackage{booktabs}
\usepackage{color}

\begin{document}

\maketitle

\begin{abstract}
The Griffin-Lim algorithm has been a seminal contribution for phase reconstruction from amplitude spectrograms, however, requiring (infinitely) high algorithmic delay. Its low-delay variant suffers in speech quality. Recent (generative) neural network methods improve on speech quality still at medium to high algorithmic delay, but often they are complex and optimized only for one specific input representation. We build upon an efficient low-delay speech vocoder and propose \texttt{EffVOC}, which supports synthesis of wideband (WB) or fullband (FB) speech from either amplitude spectrum or Mel coefficient inputs. We evaluate both input representations across multiple model sizes in a unified framework and compare against state of the art. Results show that our proposed low-delay (20 ms vs. 32 ms or more) efficient approach marks a new SOTA by achieving top-ranked subjective MOS scores (WB: 4.17/4.15, FB: 4.14/4.11) for amplitude spectrum/Mel representations, very close to ground truth.

\end{abstract}

\section{Introduction}
\label{intro}

Phase reconstruction aims to infer the phase spectrogram from amplitude-based spectral representations to subsequently reconstruct the speech waveform, enabling speech generation from representations without phase \cite{phasesummary2025}. 
Early traditional methods, such as the Griffin-Lim algorithm (\texttt{GLA}) \cite{GLA}, estimate the phase spectrogram by iteratively alternating between the time and frequency domains, enforcing consistency with the given amplitude spectrogram. Although widely used, \texttt{GLA} typically requires a high number of iterations to achieve acceptable quality, leading to high computational cost. Augmented with a momentum modification from optimization theory, an improved version of \texttt{GLA} has achieved faster convergence and improved performance \cite{fastGLA2013}. Meanwhile, adaptation of \texttt{GLA} for real-time speech applications has been realized via real-time iterative spectrogram inversion (RTISI) \cite{RTISI-LA2006}. Without further lookahead, however, this comes at the cost of reduced accuracy. A general RTISI framework capable of seamless integration with iterative algorithms has been formulated \cite{RTISI-Tal2024}, allowing straightforward adaptation and improved performance. Non-iterative algorithms \cite{SPSI2015, PGHI2017} have demonstrated efficient phase reconstruction in a single pass by directly exploiting phase-amplitude relations.
Improved performance has been achieved by leveraging the phase reconstructed by the non-iterative algorithms as an initial estimate for subsequent \texttt{GLA}-based refinement \cite{PGHI2017}.

In another direction of research, neural network-based methods have been proposed to enable faster and higher-quality waveform reconstruction either by coupling the iteration process with a neural network \cite{DeGLA2019, DeGLA2021} or by formulating a two-stage approach \cite{phase_derivative2020, phase_derivative2021} where phase derivatives are predicted in the first stage and subsequently used to reconstruct phase in the second stage. To migrate the dependency on the periodicity of estimated phase derivatives, alternative approaches conditioned on the \textit{von Mises} distribution have been introduced \cite{Gerkmann2014,phase_vonMises2018, phase2stage2021}. However, such hybrid approaches often heavily rely on traditional methods, leading to limited flexibility. Alternatively, methods that directly predict the phase spectrogram solely by a neural network have been explored \cite{nspp2023, sp_nspp2024}. Nonetheless, complex training regimes and loss functions are used to obtain reliable phase reconstruction and to ensure phase continuity. On the other hand, generative models, such as diffusion models \cite{diffphase2023} and generative adversarial networks (GANs), offer a new perspective by treating phase estimation as a generative task rather than a predictive one. Among these, GAN-based vocoders, such as \texttt{BigVGAN} \cite{bigvgan2023} and \texttt{Vocos} \cite{vocos2024}, have demonstrated high-fidelity speech reconstruction from Mel coefficient inputs and computational efficiency.

Despite the superiority of neural network-based methods, they typically rely on utterance-based reconstruction, inherently incurring (infinitely) high algorithmic delay, preventing their usage in real-time conversational applications. However, directly extending existing methods to their causal version often leads to degraded performance. To mitigate this issue, noncausal-to-causal transfer learning strategies leveraging knowledge distillation have been proposed \cite{ts2024, iwaenc2024} to regain performance. Additionally, a band-aware network \texttt{BAPEN} \cite{bapen2025} has shown to effectively capture phase patterns across frequency bands, while exhibiting minimal degradation in its causal configuration. An iterative flow-based vocoder, \texttt{MelFlow} \cite{melflow2025}, that combines generative interpolating flow matching with a causal inference scheme, has demonstrated high-performance vocoding and has achieved real-time processing on a consumer laptop GPU with a limited number of iterations. 

Meanwhile, neural speech codecs, e.g., \texttt{SoundStream} \cite{soundstream2022} and \texttt{Encodec} \cite{encodec2023}, have demonstrated superior ability in delivering high-quality speech with low delay and low bitrate while remaining architecturally efficient. 
Taking advantage of the architectural choice of \texttt{Encodec}, a low-delay speech vocoder has achieved so-far state-of-the-art (SOTA) in its class with only 20 ms of algorithmic delay \cite{vocoder_itg2025}, excelling the utterance-based \texttt{BigVGAN} and \texttt{Vocos}, as well as their causal counterparts. This method, however, only operates on Mel coefficients and synthesizes wideband speech, leaving applicability towards higher bandwidths for vocoding open. Also it is unclear, whether and how this approach can be employed for speech waveform reconstruction from the amplitude spectrum.


In this paper, we build upon the low-delay speech vocoder \cite{vocoder_itg2025} and propose \texttt{EffVOC}, which can be configured to synthesize either wideband or fullband speech, based on either amplitude spectrum input or Mel coefficient input. We employ identical window duration and frame shift by keeping the number of samples proportional to the sampling rate. We systematically investigate the impact of different spectral representations on low-delay\footnote{Note that the term \textit{delay} in this paper refers exclusively to the algorithmic delay, which is platform-independent and solely determined by the frame length of our processing scheme; delay contributions by AD/DA converters come on top. This also holds for the platform-dependent processing time, which we report in our work via the \textit{real-time factor} (RTF) on a certain platform.} speech waveform reconstruction under varying sampling rate constraints. To the best of our knowledge, we provide the first unified comparison of this kind within one reconstruction model. Finally, we analyze the effect of model depth with both inputs in the fullband setup, examining how the number of layers influences reconstruction quality. 


The rest of the paper is structured as follows. Our novel low-delay speech reconstruction model \texttt{EffVOC} is described in Section \ref{method}. The experimental setup follows in Section \ref{setups}, empirical results in Section \ref{results} and Section \ref{conclusion} provides a conclusion.

\section{Proposed Model}
\label{method}

Due to its inherent causal design and suitability for waveform reconstruction under limited algorithmic delay, we adapt the speech vocoder from \cite{vocoder_itg2025} towards our proposed low-delay speech reconstruction model \texttt{EffVOC}. In particular, we retain the efficient upsampling transformation structure to progressively increase the temporal resolution of input features to the time-domain waveform and keep recurrent layers for long-range temporal dependency modeling. Despite the simplicity, this hybrid convolution-recurrent design enables effective modeling of both local feature dynamics and temporal coherence. It can be configured to operate on Mel coefficients or amplitude spectrograms and to synthesize wideband or fullband speech.

\begin{figure}[t]
    \centering
    \label{fig:vocoder_wb}
    \includegraphics[width=0.46\textwidth]{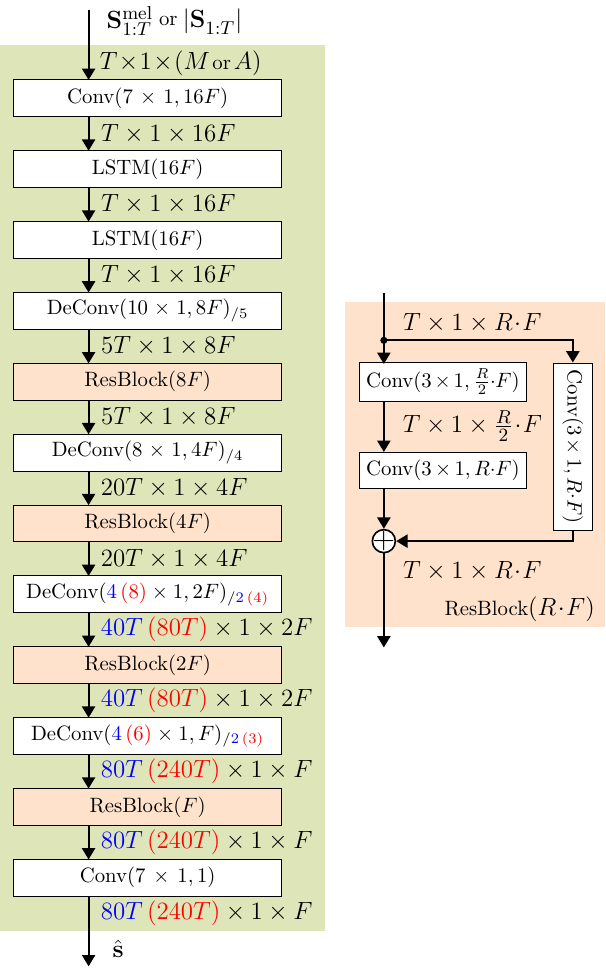}
    \vspace*{-2mm}
    \caption[toc entry]{Proposed low-delay speech reconstruction model \texttt{EffVOC} (left side) for wideband speech (black and \textcolor{blue}{blue} dimensions) and fullband speech (black and \textcolor{red}{red} dimensions). Details of the $\mathrm{ResBlock}()$ are shown on the right side.}
    \label{fig:vocoder}
\end{figure}

Details of our proposed \texttt{EffVOC} are shown in Fig.\ \ref{fig:vocoder}. The model consists of two convolutional layers, two long short-term memory (LSTM) layers, and four transposed convolutional layers, each followed by a residual block (ResBlock). Weight normalization is applied to all (transposed) convolutional layers and causal convolutions are used throughout. We use the notation $\mathrm{Conv}(h\times1, R \cdot F)$ and $\mathrm{DeConv}(h\times1, R \cdot F)_{/\textit{s}}$ to indicate a convolutional layer and transposed convolutional layer with stride \textit{s} and $R \cdot F$ kernels of size $h\times1$, respectively. The size of the feature representations before and after each layer is denoted as \textit{time axis size} $\times$ \textit{1} $\times$ \textit{number of feature maps}, with the second dimension included solely for notational consistency. The $\mathrm{ResBlock}()$ (Fig.\ \ref{fig:vocoder} right side) consists of two branches: a main branch with two consecutive convolutional layers and a bypass with a single convolutional layer. The outputs of the two branches are added. The model directly reconstructs speech waveform $\hat{\mathbf{s}}$.

To obtain the required input representations, we first segment the original time-domain speech signal $\mathbf{s}$ into overlapping frames $\mathbf{s}_{t}$ using a periodic Hann window of length $N_{w}$ and a frame shift of $N_{s}$ samples. We denote the frame index by $t \in \mathcal{T}$, where $\mathcal{T}$ represents the set of frame indices with $T = |\mathcal{T}|$ frames in an utterance. The discrete Fourier transform (DFT) of length $K$ is subsequently applied to frame $\mathbf{s}_{t}$ to obtain the spectrogram $\mathbf{S}_{t}$, from which the respective amplitude spectrogram $|\mathbf{S}_{t}|$ is extracted. Given the conjugate symmetry property of the spectrogram, we preserve only the non-redundant frequency bins to form the amplitude spectrum input, yielding $A=K/2+1$ bins in total. Built upon these, the Mel spectrogram $\mathbf{S}^{\mathrm{mel}}_{t}$ with $M$ coefficients is then obtained by applying a Mel filter bank, followed by logarithmic scaling.

The proposed \texttt{EffVOC} model architecture is mostly identical for wideband and fullband speech as shown in Fig.\ \ref{fig:vocoder}, except for the latter two transposed convolutional layers, where the strides and kernel sizes are adjusted to match the time-frequency resolution. Differences in the layer hyperparameters and the respective input and output dimensions for wideband and fullband speech are highlighted in \textcolor{blue}{blue} and \textcolor{red}{red}, respectively.

\begin{table*}[t]
    \centering
	\caption{Results of baselines and our proposed \texttt{EffVOC} model on the \textbf{VCTK test set} $\mathcal{D}^{\text{test}}_{\text{VCTK}}$ at \textbf{16 kHz} sampling rate. Models with ``*'' indicate results taken from \cite{bapen2025}. Except ground truth, the \textbf{best} and \underline{second-best} results are highlighted.}

{
\fontsize{8pt}{9pt}\selectfont
\setlength{\tabcolsep}{0.41mm}
\begin{tabular*}{16.9cm}{clcc crrccccccccc}

	\toprule

    {} & \textbf{Method} & \textbf{Input} & \textbf{\#Iter.} & \textbf{Delay} & \textbf{\#Params.} & \textbf{\#GFLOPS} & \textbf{RTF}$\big\downarrow$ & \textbf{PESQ-WB}$\big\uparrow$ & \textbf{POLQA}$\big\uparrow$ & \textbf{NISQA}$\big\uparrow$ & \textbf{LPS}$\big\uparrow$ & \textbf{ESTOI}$\big\uparrow$ & \textbf{MCD}$\big\downarrow$ & \textbf{LSD}$\big\downarrow$ & \textbf{MOS}$\big\uparrow$ \\
    \midrule
    & Ground truth & - & - & - & - & - & - & 4.64 & 4.71 & 4.27 & 1.000 & 1.00 & 0.00 & 0.00 & 4.20 \\ 
    \midrule
    & {\tt GLA} \cite{GLA} & ampl. & 100 & high & - & - & - & 4.34 & \textbf{4.55} & 4.24 & 0.971 & \underline{0.93} & 1.07 & 5.19 & 3.95 \\ 
    & {\tt RTISI-GLA} \cite{RTISI-Tal2024} & ampl. & 100 & 20 ms & - & - & - & 3.49 & 2.55 & 3.67 & 0.966 & 0.89 & 1.71 & 5.69 & 2.44 \\ 
    & {\tt RTISI-DM} \cite{RTISI-Tal2024} & ampl. & 100 & 20 ms & - & - & - & 3.57 & 2.95 & 3.79 & 0.966 & 0.92 & 1.46 & 5.28 & 2.76 \\ 
    \midrule
    & {\tt DiffPhase}* \cite{diffphase2023} & ampl. & 25 & high & 66.00 M & 6659.78 & - & \textbf{4.49} & - & - & - & - & - & - & - \\ 
    & {\tt BAPEN}* \cite{bapen2025} & ampl. & - & high & 6.82 M & 31.00 & - & \underline{4.47} & - & - & - & - & - & - & - \\ 
    & {\tt BAPEN}* \cite{bapen2025} & ampl. & - & 32 ms & 6.79 M & 31.56 & - & 4.26 & - & - & - & - & - & - & - \\ 
    \midrule

    \multirow{4}{*}{\rotatebox{90}{Ours}} & w/ F = 64 & ampl. & - & 20 ms & 27.19 M & 38.16 & 0.652 & 4.31 & \underline{4.45} & 4.19 & \textbf{0.976} & \textbf{0.94} & \textbf{0.84} & \textbf{1.80} & \textbf{4.17} \\
    & w/ F = 32 & ampl. & - & 20 ms & 7.27 M & 9.73 & 0.621 & 4.24 & 4.36 & 4.24 & 0.974 & \underline{0.93} & 0.92 & 2.18 & \textbf{4.17} \\
    & w/ F = 16 & ampl. & - & 20 ms & 2.05 M & 2.53 & 0.608 & 4.11 & 4.17 & 4.17 & 0.971 & 0.90 & 1.16 & 2.43 & 4.08 \\
    & w/ F = 8 & ampl. & - & 20 ms & 0.63 M & 0.68 & \underline{0.593} & 3.82 & 3.62 & 3.79 & 0.954 & 0.83 & 1.86 & 2.83 & 3.63 \\

    \midrule
    \midrule
    
    & {\tt MelFlow} \cite{melflow2025} & Mel & 25 & 32 ms & 27.90 M & 7050.00 & 3.594 & 4.40 & 4.32 & 4.19 & 0.973 & 0.90 & 1.17 & 2.12 & 3.95 \\ 
    & {\tt MelFlow} \cite{melflow2025} & Mel & 5 & 32 ms & 27.90 M & 1410.00 & 0.719 & 4.19 & 4.39 & \textbf{4.33} & 0.972 & 0.88 & 3.53 & 3.18 & 3.94 \\ 
    
    \midrule
    \multirow{4}{*}{\rotatebox{90}{Ours}} & w/ F = 64 & Mel & - & 20 ms & 25.92 M & 37.65 & 0.644 & 4.25 & 4.32 & 4.16 & \textbf{0.976} & \textbf{0.94} & \underline{0.88} & \underline{1.85} & 4.12 \\
    & w/ F = 32 & Mel & - & 20 ms & 6.63 M & 9.48 & 0.622 & 4.21 & 4.37 & \underline{4.27} & \underline{0.975} & \underline{0.93} & 0.94 & 2.04 & \underline{4.15} \\
    & w/ F = 16 & Mel & - & 20 ms & 1.73 M & 2.40 & 0.611 & 4.06 & 4.09 & 4.22 & 0.971 & 0.90 & 1.21 & 2.57 & 4.12 \\
    & w/ F = 8 & Mel & - & 20 ms & 0.47 M & 0.62 & \textbf{0.584} & 3.20 & 2.88 & 3.72 & 0.947 & 0.83 & 1.77 & 5.38 & 3.46 \\

	\bottomrule
    
\end{tabular*}
}
    \label{tab:ablation_16k}\vspace*{-4mm}
\end{table*}


\section{Experimental Setup}
\label{setups}

\subsection{Dataset and Preprocessing}
\label{dataset}

We evaluate the performance of our proposed low-delay speech reconstruction model \texttt{EffVOC} on a multi-speaker English-language VCTK dataset \cite{VCTK}. All recordings have an original sampling rate of 48 kHz. To ensure a fair comparison, we use the same test set $\mathcal{D}^{\text{test}}_{\text{VCTK}}$ as \texttt{BAPEN} \cite{bapen2025}, which consists of recordings from the last 8 speakers. Training and validation sets are configured with the remaining recordings. Our training set $\mathcal{D}^{\text{train}}_{\text{VCTK}}$ consists of approximately 35.8 hours of recordings from 93 speakers. The validation set $\mathcal{D}^{\text{val}}_{\text{VCTK}}$ is formed with recordings from the remaining speakers and has a total duration of 2.5 hours. We re-sample all recordings from three splits to 16 kHz for the wideband experiments. 

As mentioned in the introduction, we use fixed window and frame shift durations for both wideband and fullband experiments. Specifically, a periodic Hann window of 20 ms with a frame shift of 5 ms (75\% frame overlap) is employed in all experiments, resulting in a window length and frame shift pair $(N_{w}, N_{s}) = (320, 80)$ for wideband and $(960, 240)$ for fullband, respectively. The corresponding DFT sizes are $K=512$ and $K=1024$, respectively. Our Mel filterbank consists of $M=80$ Mel filters for both sampling rates. The algorithmic delay of our proposed \texttt{EffVOC} is 20 ms.

\subsection{Experimental Settings and Training}
\label{training}

\textbf{Model configurations:} We investigate the effect of amplitude spectrum and Mel coefficient inputs using four different model sizes in the wideband experiments by setting $F \in \{64, 32, 16, 8\}$. On the other hand, we set $F=32$ for the fullband experiments and examine the impact of model depth with three different stride setups, i.e., $\{5,4,4,3\}$ as in Fig.\ \ref{fig:vocoder}, $\{5,4,3,2,2\}$, and $\{5,3,2,2,2,2\}$, respectively. To accommodate the latter two stride setups, the number of transposed convolutional layers is increased to five and six, respectively, each followed by a residual block. Following the same design in Fig.\ \ref{fig:vocoder}, the number of kernels in each transposed convolutional layer is halved w.r.t.\ the previous layer.


\noindent\textbf{Training setup:} We follow the training procedure \cite{vocoder_itg2025} to train our proposed \texttt{EffVOC}. An AdamW optimizer with an initial learning rate of $1 \cdot 10^{-4}$ and a batch size of 32 is used. Other configurations, including the discriminator setup, the optimizer hyperparameter, the learning rate scheduler, and the training losses, follow the official implementation of \texttt{BigVGAN} \cite{bigvgan2023}. All models are trained with 1M steps.

\noindent\textbf{Baseline models:} We compare our proposed \texttt{EffVOC} with both traditional algorithms, i.e., \texttt{GLA} \cite{GLA}, \texttt{RTISI-GLA} \cite{RTISI-Tal2024}, a low-delay difference map version (\texttt{RTISI-DM}) \cite{RTISI-Tal2024}, and neural network-based models, i.e., \texttt{DiffPhase} \cite{diffphase2023}, \texttt{BAPEN} \cite{bapen2025}, and \texttt{MelFlow} \cite{melflow2025}. The same pair of window length and frame shift detailed in Section \ref{dataset} is used for \texttt{GLA}, \texttt{RTISI-GLA}, and \texttt{RTISI-DM}. While \texttt{MelFlow} is trained with 1M steps on the same dataset as our methods, it follows its original configuration from \cite{melflow2025} with a 32 ms periodic Hann window and 16 ms frame shift (50\% frame overlap). Results of \texttt{DiffPhase} and \texttt{BAPEN} are directly taken from \cite{bapen2025}.


\begin{table*}[t]
    \centering
	\caption{Results of baselines and our proposed \texttt{EffVOC} model with $F=32$ on the \textbf{VCTK test set} $\mathcal{D}^{\text{test}}_{\text{VCTK}}$ at \textbf{48 kHz} sampling rate. Model with ``*'' indicates result taken from \cite{bapen2025}. Except ground truth, the \textbf{best} and \underline{second-best} results are highlighted.}\vspace*{-2mm}

{
\fontsize{8pt}{9pt}\selectfont
\setlength{\tabcolsep}{0.7mm}
\begin{tabular*}{16.9cm}{clcc crrcccccccc}

	\toprule

    {} & \textbf{Method} & \textbf{Input} & \textbf{\# Iter.} & \textbf{Delay} & \textbf{\# Params.} & \textbf{\# GFLOPS} & \textbf{RTF}$\big\downarrow$ & \textbf{PESQ-WB}$\big\uparrow$ & \textbf{POLQA}$\big\uparrow$ & \textbf{NISQA}$\big\uparrow$ & \textbf{ESTOI}$\big\uparrow$ & \textbf{MCD}$\big\downarrow$ & \textbf{LSD}$\big\downarrow$ & \textbf{MOS}$\big\uparrow$ \\
    \midrule
    & Ground truth & - & - & - & - & - & - & 4.64 & 4.67 & 4.46 & 1.00 & 0.00 & 0.00 & 4.15 \\ 
    \midrule
    & {\tt GLA} \cite{GLA} & ampl. & 100 & high & - & - & - & \underline{4.26} & 4.30 & 4.43 & \textbf{0.93} & 1.62 & 5.25 & 3.94 \\ 
    & {\tt RTISI-GLA} \cite{RTISI-Tal2024} & ampl. & 100 & 20 ms & - & - & - & 3.48 & 2.51 & 3.87 & 0.89 & 1.98 & 5.42 & 2.62 \\ 
    & {\tt RTISI-DM} \cite{RTISI-Tal2024} & ampl. & 100 & 20 ms & - & - & - & 3.56 & 2.91 & 3.95 & 0.92 & 1.83 & 2.14 & 2.83 \\ 
    & {\tt BAPEN}* \cite{bapen2025} & ampl. & - & high & 6.82 M & 31.00 & - & \textbf{4.43} & - & - & - & - & - & - \\ 
    
    \midrule
    
    \multirow{3}{*}{\rotatebox{90}{Ours}} & w/ \{5,4,4,3\} & ampl. & - & 20 ms & 8.22 M & 13.39 & \textbf{0.630} & 4.22 & \textbf{4.33} & 4.40 & \textbf{0.93} & \textbf{1.11} & \textbf{2.01} & \textbf{4.14} \\
    & w/ \{5,4,3,2,2\} & ampl. & - & 20 ms & 8.20 M & 11.53 & 0.749 & 4.22 & \textbf{4.33} & 4.41 & \underline{0.92} & 1.18 & \underline{2.07} & \underline{4.12} \\
    & w/ \{5,3,2,2,2,2\} & ampl. & - & 20 ms & 8.12 M & 8.90 & 0.878 & 4.22 & \underline{4.31} & 4.40 & \textbf{0.93} & \underline{1.14} & 2.27 & 4.11 \\
    
    \midrule
    \midrule

    & {\tt MelFlow} \cite{melflow2025} & Mel & 25 & 32 ms & 28.10 M & 21130.00 & 3.856 & 4.22 & 4.14 & 4.38 & 0.89 & 2.16 & 3.14 & 3.89 \\ 
    & {\tt MelFlow} \cite{melflow2025} & Mel & 5 & 32 ms & 28.10 M & 4226.00 & 0.771 & 4.10 & 4.19 & \underline{4.44} & 0.88 & 4.81 & 4.41 & 3.85 \\ 
    
    \midrule
    \multirow{3}{*}{\rotatebox{90}{Ours}} & w/ \{5,4,4,3\} & Mel & - & 20 ms & 6.67 M & 12.77 & \underline{0.632} & 4.08 & 4.05 & \textbf{4.45} & \underline{0.92} & 1.31 & 2.33 & 4.10 \\
    & w/ \{5,4,3,2,2\} & Mel & - & 20 ms & 6.65 M & 10.91 & 0.748 & 4.07 & 4.10 & \underline{4.44} & \underline{0.92} & 1.31 & 2.37 & 4.10 \\
    & w/ \{5,3,2,2,2,2\} & Mel & - & 20 ms & 6.57 M & 8.28 & 0.864 & 4.04 & 4.10 & \textbf{4.45} & \underline{0.92} & 1.36 & 2.50 & 4.11 \\
    
    
	\bottomrule
    
\end{tabular*}
}
    \label{tab:ablation_48k}
\end{table*}


\subsection{Evaluation Metrics}
\label{metrics}

\textbf{Instrumental metrics:} Several metrics are employed for evaluation. Five intrusive metrics are selected, including the perceptual evaluation of speech quality (PESQ-WB) according to ITU-T Recommendation P.826.2 \cite{PESQ}, the perceptual objective listening quality analysis (POLQA) \cite{POLQA}, extended short-time objective intelligibility (ESTOI) \cite{estoi}, the Mel-cepstral distance (MCD) \cite{MCD}, and the log-spectral distance (LSD) \cite{lsd}. Moreover, the non-intrusive metric NISQA \cite{nisqa} and the Levenshtein phone similarity (LPS) \cite{lpd2023} are used. Note that LPS$\,$=$\,$1$\,$-$\,$LPD, with the Levenshtein phoneme distance (LPD) taken from \cite{lpd2023}. We also measure the real-time factor (RTF) on a single \texttt{NVIDIA A100} GPU device as the ratio of the time to process one single frame divided by the frame shift. An RTF $\leq 1$ signifies real-time capability. As mentioned earlier, the frame shift for our \texttt{EffVOC} is 5 ms, whereas \texttt{MelFlow} uses a 16 ms frame shift.

\noindent\textbf{Subjective metrics:} In addition, we perform ITU-T P.808 \cite{p808} subjective listening tests, receiving absolute category rating mean opinion scores (MOS) via crowdsourcing on a 7\% subset of $\mathcal{D}^{\text{test}}_{\text{VCTK}}$ (in total 160 files, with the same gender distribution). Following the URGENT 2024 Challenge approach \cite{urgent_challenge2024}, this subset was carefully designed using constrained pseudo-random file selection to preserve the rank order of all instrumental metrics, thereby ensuring the same distribution as the full test set. For each evaluated model, the MOS was computed as the mean over 160 files, each rated by 8 different listeners. All listeners passed a strict check according to ITU-T P.808 and received compensation.

\section{Results and Discussion}
\label{results}

Table \ref{tab:ablation_16k} compares our proposed \texttt{EffVOC} against all baselines on $\mathcal{D}^{\text{test}}_{\text{VCTK}}$ in a \textit{wideband} setup with 16 kHz sampling rate. Model size is measured in terms of trainable parameter count (\# Params.) and complexity as Giga floating-point operations per second (\# GFLOPS). 

The upper part of Table \ref{tab:ablation_16k} shows models with \textit{amplitude spectrum inputs}. We observe an overall high subjective quality of \texttt{GLA} \cite{GLA}, being top in POLQA. Other utterance-based neural networks \cite{diffphase2023, bapen2025} show top PESQ-WB scores. The low-delay (i.e., 20 ms) version of \texttt{GLA} and \texttt{DM} \cite{RTISI-Tal2024} severely fall behind in all metrics, motivating the search for good quality low-delay methods. The larger version of our proposed model ($F=64$) shows good PESQ-WB and POLQA scores, while being overall strongest in speech intelligibility (LPS, ESTOI) and in signal fidelity (MCD, LSD) among all reported methods. Comparing our amplitude-based model ($F=64$) to \texttt{BAPEN} \cite{bapen2025} with 32 ms algorithmic delay, we reach a slightly higher PESQ-WB (4.31 vs.\ 4.26) with a larger model at only slightly higher complexity, but requiring only 20 ms of algorithmic delay. Using a similar model size ($\sim$ 7 M, $F=32$), a comparable PESQ-WB (4.24 vs.\ 4.26) is reached at much lower computational demand.

Models with \textit{Mel coefficient inputs} are shown in the lower part of Table \ref{tab:ablation_16k}. \texttt{MelFlow} \cite{melflow2025}, a Mel representation baseline with 32 ms algorithmic delay, is reported both in a high-performance variant (25 iterations, very good PESQ-WB score) and in a single-GPU real-time capable streaming variant (5 iterations, good RTF of 0.719). Our Mel-based model ($F=64$) delivers still a good PESQ-WB while being top in speech intelligibility (LPS, ESTOI) and overall second-best in signal fidelity (MCD, LSD). Our $F=32$ variant provides roughly the same speech quality scores and LPS as \texttt{MelFlow} with 5 iterations, but with lower algorithmic delay, much smaller model size, and requiring two orders of magnitude less computations (GFLOPS). On top, we reach better ESTOI and signal fidelity (MCD, LSD). Also, the RTFs of our $F=64$ and $F=32$ models are slightly better than \texttt{MelFlow} with 5 iterations.
Note that a higher frame overlap and a shorter analysis frame are known to ease the reconstruction problem through higher time-frequency redundancy, in particular when mapping fewer DFT bins to the same number of $M=80$ Mel channels. The choices we make for \texttt{EffVOC} (20 ms frames, 75\% overlap) hence lead to a higher frame rate and higher computational cost than 50\% overlap, but allow for a lower delay and higher reconstruction fidelity than longer frames and less overlap.

Among our proposed models, we observe that for $F=64$ and $F=32$, amplitude and Mel representations have comparable performance. Only for smaller network sizes, the amplitude input representation delivers a significantly better quality. 

\textit{Concerning MOS scores, we clearly see our proposed networks ahead of all others, with our low-delay $F=64$ and $F=32$ amplitude-based models being in top position, closely followed by our Mel-based network with $F=32$; both within an only 0.05 MOS margin towards ground truth.}

In Table \ref{tab:ablation_48k}, we evaluate the performance of our proposed \texttt{EffVOC} with $F=32$ on $\mathcal{D}^{\text{test}}_{\text{VCTK}}$ in a \textit{fullband} setup with 48 kHz sampling rate. Smae as in Table \ref{tab:ablation_16k}, models are grouped by the respective input representation. 

For \textit{amplitude spectrum inputs}, again, the (infinite) delay \texttt{BAPEN} method \cite{bapen2025} is very strong in PESQ-WB, but they don't report on the fullband POLQA metric, where our methods are leading overall, even excelling \texttt{GLA} \cite{GLA}. Again, our methods are best in intelligibility (ESTOI) and signal fidelity (MCD, LSD). 

Models with \textit{Mel coefficient inputs} are shown in the lower part of Table \ref{tab:ablation_48k}. \texttt{MelFlow} (32 ms algorithmic delay) \cite{melflow2025} in both variants, again, is best among Mel input methods for PESQ-WB and POLQA, but with large model size and very high complexity. Still, the streaming \texttt{MelFlow} (5 iterations) reaches real-time capability (RTF of 0.771). Interestingly, RTFs of our methods are in a similar range. Our 20 ms low-delay approaches are again stronger both in intelligibility and signal fidelity. Across our proposed models, we observe that amplitude input models consistently outperform Mel input models w.r.t.\ PESQ-WB and POLQA and achieve better signal fidelity.


\textit{Concerning MOS scores for fullband speech, all of our models outperform any of the baselines.} We find it particularly interesting that even our smallest proposed model (Mel coefficient input, 6.57 M) reaches a competitive MOS of 4.11, excelling all baseline methods and being less than 0.05 MOS points behind ground truth, \textit{thereby setting a new SOTA for fullband speech synthesis from Mel coefficients.}

\section{Conclusions}
\label{conclusion}

We investigated low-delay speech waveform reconstruction using amplitude spectrum and Mel coefficient inputs in a unified speech vocoder framework. Among the neural network approaches, our proposed real-time capable \texttt{EffVOC} models are small, low-complex, and of low algorithmic delay (20 ms). For \textit{wideband} speech, our \texttt{EffVOC} demonstrates new state of the art (SOTA) subjective MOS scores. Our smallest proposed \textit{fullband} \texttt{EffVOC} model (Mel representation, 6.57 M) reaches a competitive MOS of 4.11, excelling all baselines and being less than 0.05 MOS points behind ground truth, thereby setting a new SOTA for fullband speech synthesis from Mel coefficients.

\clearpage
\section{Use of Generative AI Disclosure}
This manuscript was written without the use of Generative AI.

\bibliographystyle{IEEEtran}
\bibliography{mybib}

\end{document}